\documentclass[aps,prb,twocolumn,floatfix,superscriptaddress]{revtex4}

\usepackage{graphicx,graphics}

\begin{document}

\title{Intrinsic magnetism at silicon surfaces}
\thanks{The final version of this draft preprint was published as Nat.~Commun.~{\bf 1:58} (2010) and is available at http://dx.doi.org/10.1038/ncomms1056.}
\author{Steven C. Erwin}
\affiliation{Center for Computational Materials Science, Naval Research Laboratory, Washington, DC 20375, USA}
\author{F. J. Himpsel}
\affiliation{Department of Physics, University of Wisconsin-Madison, Madison, WI 53706, USA}

\date{April 23, 2010 (submitted version)}

\begin{abstract}
  It has been a long-standing goal to create magnetism in a
  nonmagnetic material by manipulating its structure at the nanometer
  scale.  This idea may be realized in graphitic carbon: evidence
  suggests magnetic states at the edges of graphene ribbons and at
  grain boundaries in graphite. Such phenomena have long been regarded
  as unlikely in silicon because there is no graphitic bulk
  phase. Here we show theoretically that intrinsic magnetism indeed
  exists in a class of silicon surfaces whose step edges have a
  nanoscale graphitic structure. This magnetism is intimately
  connected to recent observations, including the coexistence of
  double- and triple-period distortions and the absence of edge states
  in photoemission. Magnetism in silicon may ultimately provide a
  path toward spin-based logic and storage at the atomic level.
\end{abstract}


\maketitle

Individual atoms with an odd number of electrons exhibit a magnetic
moment from the spin of the unpaired electron.  Elements with an even
number of electrons, such as carbon and silicon, can also reveal
unpaired electrons when covalent bonds are broken.  In defective or
disordered group-IV solids the single electrons that occupy dangling
bonds can be detected in magnetic resonance experiments, but the lack
of long-range structural order precludes a magnetically
ordered state.

Bonds are also broken at the surfaces and interfaces of solids, but
covalent materials usually reconstruct to eliminate dangling bonds or
doubly occupy them with electrons of opposite spin. Exceptions to this
behavior provide fertile ground for magnetism in nonmagnetic
materials.  Two widely discussed examples are based on graphitic
carbon: highly oriented pyrolytic graphite exhibits ferromagnetic
order arising from two-dimensional arrays of defect spins at grain
boundaries,\cite{cervenka_nature_phys_2009a} and nanoscale graphene
ribbons are predicted to have ferromagnetically ordered edge
states.\cite{lee_phys_rev_b_2005a} 

The edges of graphene ribbons are theoretically appealing but far from
being realized experimentally with the required atomic smoothness.
Other one-dimensional systems such as nanowires
\cite{zabala_phys_rev_lett_1998a} and step edges provide more
favorable conditions for long-range structural order and thus are 
promising platforms for devices requiring ordered arrays of spins.
Indeed, atomically smooth steps with structural order extending to
micrometers can be produced by self-assembly at vicinal silicon
surfaces.  Some of these surfaces naturally form a graphitic silicon
ribbon at the step edge.  Here we predict, using density-functional
theory (DFT, see Methods), that the electronic ground state of these
silicon surfaces is magnetic. Magnetism in silicon systems having no
magnetic elements is quite unexpected. Indeed, none of these systems has yet
been investigated for magnetism experimentally, but we show below that
existing data already support it indirectly.

\begin{figure*}
\includegraphics[width=16cm]{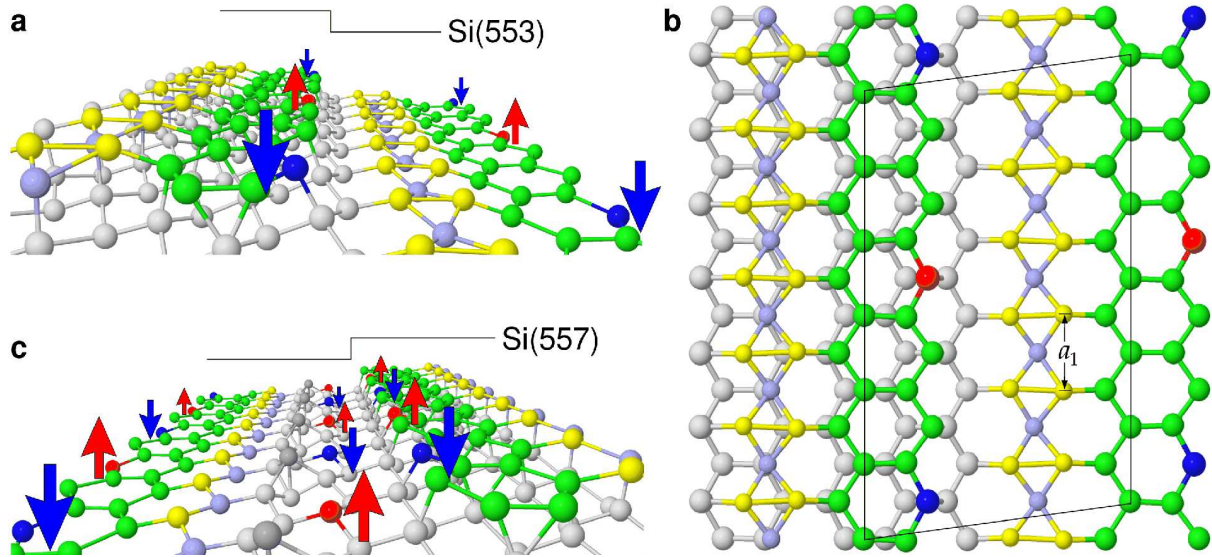}
\caption{{\bf Ground state structure and lowest energy spin
configuration of two magnetic silicon
surfaces.} {\bf a--b,} Si(553)-Au in its antiferromagnetic
ground state.  Yellow atoms are
Au, all others are Si.  Each terrace contains
a Au double row and a graphitic Si honeycomb chain (green) at
the step edge.   Every third Si (red, blue) at the step has a spin
magnetic moment of one Bohr magneton ($S=1/2$, arrows) from the
complete polarization of the electron occupying the dangling-bond 
orbital. The sign of the polarization alternates along the step.
The periodicity along the step is tripled by a small downward
displacement of the magnetic atoms. The periodicity along the two Au 
rows is doubled, with alternating short and long bonds $a_1$.
The full 1$\times$6 unit cell is outlined.
{\bf c,} Si(557)-Au in its antiferromagnetic ground state. The step
direction of Si(557) is opposite to that of Si(553) as shown by the 
schematic outlines. Nevertheless this surface is also magnetic:
every second Si (red, blue) at the step has a magnetic moment
of one Bohr magneton, as do all of the Si restatoms (blue, red) on the
terrace.
\label{model}}
\end{figure*}

Our discussion focuses on the Si(553)-Au and Si(557)-Au surfaces,
shown in Fig.~1. These are stepped surfaces stabilized by a fraction of a monolayer of
Au. Crystallographically they can be considered miscut from the flat
Si(111) surface by angles of 12.3$^\circ$ and 9.4$^\circ$ away from
and toward, respectively, the (001) orientation. They are members of a
family of Au-induced vicinal silicon surfaces intermediate between
(111) and (001) containing several structural motifs in common.
\cite{crain_phys_rev_b_2004a} These commonalities suggest that
magnetism in silicon may be widespread.

The critical structural motif for magnetism in these systems is a
nanoscale honeycomb strip of graphitic silicon, just a single hexagon
wide, that forms the edge of the step.\cite{erwin_phys_rev_lett_1998a} This honeycomb chain was first
observed in reconstructions of flat Si(111) induced by submonolayer
coverages of alkali, alkaline-earth, and some rare-earth metals, as
well as Ag and Au.\cite{barke_solid_state_comm_2007a} In these flat
surfaces the orbitals of the outer honeycomb atoms form filled bands,
either by accepting electrons from the adsorbates or by forming
partially covalent bonds. At a step edge a new possibility arises in
which these orbitals are singly occupied, fully spin polarized, and
magnetically ordered.

The two surfaces studied here have step edges with opposite
orientation. The DFT ground state (see Methods) depicted in Fig.~1a for Si(553)-Au
shows every third Si atom at the step edge to be completely spin
polarized while all other atoms are negligibly polarized. On
Si(557)-Au (Fig.~1c) every second Si atom at the step is fully
polarized as are all the terrace restatoms.  In both systems the
interaction between neighboring spins along a step is antiferromagnetic.

These predictions depend on an accurate knowledge of the atomic
structure of each surface. The model of Si(557)-Au shown in Fig.~1c is
well-established from DFT calculations
\cite{sanchez-portal_phys_rev_lett_2004a,crain_phys_rev_b_2004a} and X-ray diffraction.\cite{robinson_phys_rev_lett_2002a}  The energy bands measured in
Si(557)-Au agree well with the calculated bands arising from the Au
chains. A small band splitting\cite{segovia_nature_1999a,losio_phys_rev_lett_2001a,altmann_phys_rev_b_2001a,ahn_phys_rev_lett_2003a} originates from the spin-orbit interaction
of the Au atoms.\cite{sanchez-portal_phys_rev_lett_2004a,barke_phys_rev_lett_2006a}
In this ``Rashba effect''--- which is unrelated to the magnetism we
predict here---the two split bands have a spin polarization of 100\%
at any given momentum in reciprocal space, but this vanishes when
integrated over all momenta and leaves no net spin polarization in real space.

Existing models for the structure of the Si(553)-Au surface do not
reproduce the experimental band structure.\cite{riikonen_surf_sci_2006a,riikonen_phys_rev_b_2008a}  We demonstrate below that
the model proposed in Fig.~1a correctly predicts not only the band
structure, but also the previously unexplained 
coexistence of two periodicities observed in scanning tunneling microscopy
(STM).  We also show that these predictions are inextricably linked
with magnetic order.  Their experimental confirmation constitutes
indirect but compelling evidence for the existence of magnetism as well.

\begin{figure}
\includegraphics[width=8cm]{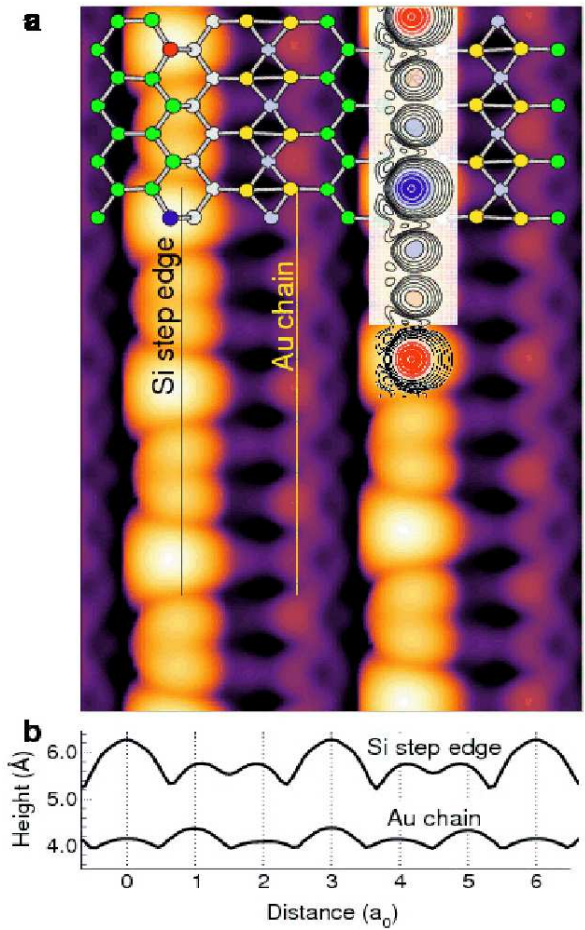}
\caption{
{\bf Scanning tunneling microscopy of Si(553)-Au.} {\bf a,} Theoretical STM topography
showing the tip height at constant current for tunneling into empty
surface states at bias voltage
$+$0.5 V. Inset: Electron spin density 1 \AA~above the step
edge. Red and blue contours represent positive and negative polarization,
indicating antiferromagnetic alignment of the red and blue
Si atoms the step edge. Contours are separated by a factor of two. 
{\bf b,} Line scans at the locations marked in {\bf a}, showing the
coexisting 3$a_0$ and 2$a_0$ periodicity along the Si step edge and
Au chain, respectively.
\label{stm}}
\end{figure}

The Si(553)-Au surface was first observed in
Ref.~\onlinecite{crain_phys_rev_lett_2003a} and its unusual one-dimensional
electronic properties have since been extensively studied. Two
experimental findings are especially striking. (1) STM images acquired
at room temperature show alternating bright and dim rows, each with
the periodicity of the Si surface lattice constant
$a_0$.\cite{crain_science_2005a,ahn_phys_rev_lett_2005a,snijders_phys_rev_lett_2006a,crain_phys_rev_lett_2006a}
At temperatures below 50 K these rows separately develop higher-order
periodicity: a tripled period (3$a_0$) for the bright rows and a
doubled period (2$a_0$) for the dim rows. These effects have variously
been attributed to periodic lattice distortions
\cite{ahn_phys_rev_lett_2005a} and to charge-density
waves,\cite{snijders_phys_rev_lett_2006a} but their true microscopic
origin has remained elusive.  (2) Angle-resolved photoemission spectra
(ARPES) obtained at 160--220 K reveal three metallic bands, all with
their minima at the zone boundary ZB$_\parallel$ of the 1$\times$1
Brillouin zone.\cite{crain_phys_rev_lett_2003a} The two lower bands
are slightly split by a small momentum splitting \mbox{$\delta
  k_F=0.05$ \AA$^{-1}$},\cite{barke_phys_rev_lett_2006a} and cross the
Fermi level about halfway to the zone center.  Detailed analysis of
this crossing shows that the splitting is due to the spin-orbit
interaction.\cite{barke_phys_rev_lett_2006a} The upper band crosses
the Fermi level about one-fourth of the way to the zone center without
a detectable splitting.

All of these observations are correctly predicted by the proposed
model of Si(553)-Au. Figure 2 shows the theoretical simulated STM
image for the model in Fig.~1a in its antiferromagnetic ground state.
The bright row arises from Si atoms at the step edge and indeed has
tripled periodicity 3$a_0$. The highest peaks within this row are from
the spin-polarized atoms. This appears counterintuitive since these
atoms are actually 0.3 \AA~{\em below} the height of their
non-polarized neighbors. Moreover, the dangling-bond orbital on this
atom is occupied by a single electron rather than two.  But the
average energy of the occupied spin-polarized orbital is energetically
closer to the Fermi level and as a result the topography appears higher in 
filled-state tunneling.

The dim second row visible in Fig.~2a arises from the right-hand leg
of the ladder formed by two rows of Au atoms. Its doubled periodicity
2$a_0$ is due to the dimerization created when the rungs of this
ladder rotate with alternating signs along a row.  A similar
dimerization occurs in a closely related system,
Si(111)-(5$\times$2)-Au, where it it is driven by electron doping from
Si adatoms and the consequent formation of a surface band gap.\cite{erwin_phys_rev_b_2009a} For Si(111)-(5$\times$2)-Au the
dimerization parameter $d=(a_1-a_0)/a_0$, where $a_0$ is
the surface lattice constant and $a_1$ is defined in Fig.~1b, has the
value $d=0.14$, much larger than the value $d=0.04$ for Si(553)-Au.
But even this small dimerization is sufficient to produce a doubled
periodicity that is well-resolved in the theoretical STM image.

The spin polarization of the electron density, shown in the inset to
Fig.~2a, is strongly localized at every third Si step-edge atom.
Integrating this spin density shows that the red and blue step-edge
atoms each have a spin moment of 1 Bohr magneton, equivalent to a
fully polarized electron with spin $S=1/2$. The direction of the spin
alternates along the step in this antiferromagnetic ground state.
Moments on other atoms along the step are smaller by an order of
magnitude.

The connection between magnetism and the periodicity of the rows is
simple and direct. If the spin polarization is constrained to be zero
then the DFT equilibrium geometry of the model changes: both the
tripled periodicity along the Si rows and the doubled periodicity
along the Au rows are completely eliminated. In this sense the
experimental observation of triple and double periodicities
constitutes strong evidence for a spin-polarized ground state.

Magnetism also provides the key to understanding a long-standing
puzzle presented by ARPES data for Si(553)-Au. The orbitals at the
step edges of a vicinal surface would normally give rise to sharp
electronic states near the Fermi level, as is indeed the case for previously proposed
models of Si(553)-Au.\cite{riikonen_surf_sci_2006a,riikonen_phys_rev_b_2008a}
But these are not observed in the ARPES data. We demonstrate below that
spin polarization at the step edge provides the resolution to this puzzle.

\begin{figure*}
\includegraphics[width=16cm]{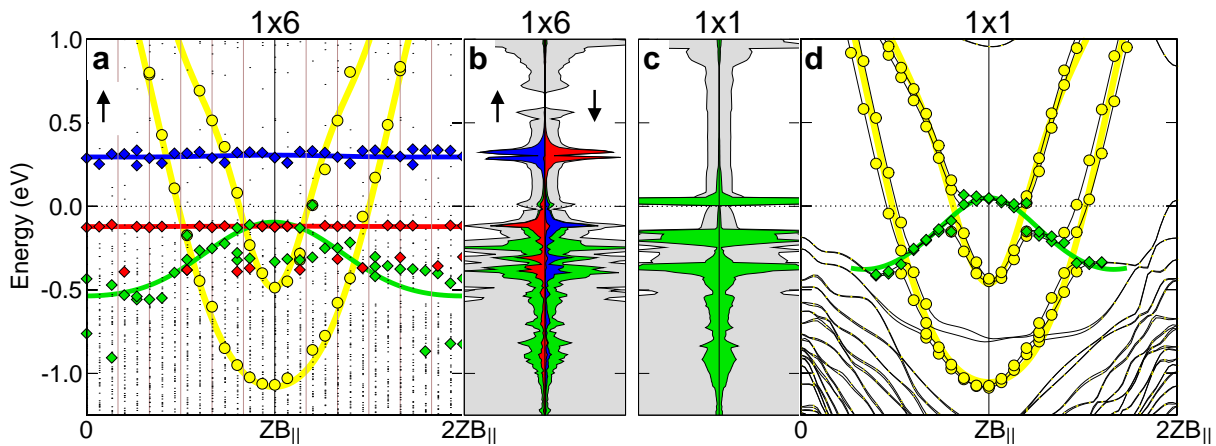}
\caption{{\bf Electronic band structure and density of states for
  Si(553)-Au.} {\bf a--b,} Bands and DOS for the ground-state
  antiferromagnetic structure of Fig.~1a. The full 1$\times$6 bands are
  shown unfolded and parallel to the chain direction (ZB$_\parallel$ is the
  boundary of the 1$\times$1 Brillouin zone) along the line
  halfway to the orthogonal zone boundary ZB$_\perp$.  Red and
  blue diamonds are spin-polarized states from red and blue Si
  step-edge atoms in Fig.~1a. Only spin-up states are
  shown here---antiferromagnetic order implies the spin-down
  states are equivalent but with red and blue reversed, as seen in the DOS.  Green
  diamonds are non-polarized states from green Si step-edge
  atoms.  Yellow circles are bonding (lower) and antibonding
  (upper) states from the Au chains.  Curves show schematic energy bands
  without hybridization.
 {\bf c--d,} Bands and DOS
  for the same model when it is constrained to 1$\times$1 periodicity,
  for which the ground state is
  not spin-polarized. Spin-orbit coupling, included in the bands
  but not in the DOS, splits each of the two Au chain states but not
  the Si step-edge state.
  \label{bands+dos}}
\end{figure*}

Figure 3a shows the theoretical bands for Si(553)-Au
unfolded into the extended Brillouin zone, where they can be easily
compared to those obtained from ARPES spectra.  Since the theoretical
ground state is antiferromagnetic, one need only analyze the bands
from a single spin channel.  The symbols and curves identify, using
the colors from Fig.~1, the atoms that dominate the corresponding
wavefunctions.  States arising from the magnetic atoms experience an
exchange splitting of about 0.5 eV.  As a result, for spin-up states
the red step-edge band is singly occupied and the blue step-edge band
is empty. For spin-down states this situation is reversed, as seen in
the density of states, Fig.~3b.

While these magnetic step-edge states are responsible for the
1$\times$6 structural and magnetic periodicity of Si(553)-Au, the
remaining features of the band structure have simple 1$\times$1
nonmagnetic character. This is evident from comparing the 1$\times$6
bands to those in Fig.~3d for a simpler 1$\times$1 structure identical
to our model but constrained to unit periodicity $a_0$ and thus having
no spin polarization. In this 1$\times$1 structure there is just a
single type of step atom and therefore only one corresponding band
(green). The remnant of this nonmagnetic band is also visible in the
full 1$\times$6 bands, although there its spectral weight at a given
momentum is spread out by partial hybridization with its magnetic
counterparts.

Both the 1$\times$6 and 1$\times$1 structures give rise to several
metallic bands (yellow) with their minima at the zone boundary
ZB$_\parallel$. These are the bands detected in ARPES. In the
1$\times$6 case there are a total of four bands in the two spin
channels, but the spin-up and spin-down bands are nearly degenerate.
In the nonmagnetic 1$\times$1 case there are also four bands but here
the degeneracy is broken by the spin-orbit
interaction,\cite{koelling_j_physics_c_solid_state_phys_1977a} which
for simplicity was not included in the 1$\times$6
calculation.\cite{notespinorbit} The resulting momentum splitting
\mbox{$\delta k_F=0.04$ \AA$^{-1}$} of the lower band is in excellent
agreement with the experimental splitting.  The splitting of the upper
band is less than half this value and consequently is not resolved in
the ARPES data. The predicted Fermi-level crossings of these bands are
in qualitative agreement with the measured values: the lower doublet
crosses the Fermi level halfway to the zone center, and the upper
doublet about one-fifth of the way.

One feature of the nonmagnetic 1$\times$1 band structure is {\em not}
observed in the ARPES data: the green step-edge band that just crosses
the Fermi level in Fig.~3d. Magnetism provides a simple explanation
for the apparent absence of this band. Above a temperature scale set
by the energy gained from spin polarization, thermal fluctuations
between polarized and unpolarized electron states will broaden the
step-edge bands.  The magnitude of the broadening is roughly the exchange
splitting, 0.5 eV. The resulting reduction in intensity makes the
step-edge bands essentially invisible compared to the sharp Au bands,
which are not affected by the spin polarization at the step edge. In
this sense the experimental absence of a sharp step-edge band
constitutes additional evidence for a spin-polarized ground state.

We now briefly summarize our predictions for magnetism in Si(557)-Au,
which are qualitatively similar to those for Si(553)-Au, and defer a
detailed discussion. Despite having a more complicated arrangement of
spins, the surface has a simpler 1$\times$2 periodicity which persists
even if spin polarization is suppressed. Magnetism leads to quantitative
but not qualitative structural changes: a smaller downward
displacement of the magnetic step-edge atoms and a smaller upward
displacement of the magnetic restatoms.  Therefore we do not predict
any higher-order periodicity to mark the onset of magnetism at low
temperature.

Earlier theoretical studies of Si(557)-Au included spin-orbit coupling
but not spin polarization.\cite{sanchez-portal_phys_rev_lett_2004a}
The resulting prediction of momentum-split Au bands were in good
agreement with experiment. But these studies also predicted Si
step-edge and restatom bands not seen in ARPES.\cite{crain_phys_rev_b_2004a,losio_phys_rev_lett_2001a,ahn_phys_rev_lett_2003a} This discrepancy is resolved in the
magnetic ground state: the exchange splitting pushes the two occupied
spin bands down to 0.4--0.5 eV below the Fermi level, and the two
unoccupied spin bands up to 0.1--0.2 eV above the Fermi level. 
This exchange splitting is similar to the splitting in
Si(553)-Au. Thermal fluctuations between polarized and unpolarized
states will likewise broaden these step-edge and restatom bands,
rendering them invisible to ARPES while leaving the momentum-split Au
bands unaffected.

We anticipate that spin-polarized states in silicon are not limited to Si(553)-Au
and Si(557)-Au: the key structural element---a graphitic step
edge---is known to exist on other stepped silicon surfaces as
well.\cite{crain_phys_rev_b_2004a} Direct experimental tests of our
predictions using, for example, spin-polarized STM
\cite{bode_rep_prog_phys_2003a,wiesendanger_rev_mod_phys_2009a} or
spin-polarized photoemission \cite{osterwalder2006a} will be an
important next step, and will provide important information about the
nature of the magnetic ordering.

To set the stage for such tests it is helpful to estimate the
magnitude of the spin interactions within a simple nearest-neighbor
Heisenberg Hamiltonian.  From DFT total-energy calculations of
different spin configurations on Si(553)-Au we find antiferromagnetic
coupling ($J_\parallel=15$ meV) along the steps and weaker,
ferromagnetic coupling ($J_\perp=-0.3$ meV) across the steps. We also
find that the magnitude, and even the sign, of these couplings can
be changed by doping electron or holes into the Si(553)-Au surface
states. It is well-established that surface states on the closely
related Si(111)-Au surface can be electron-doped by adsorbates
(e.g.~silicon adatoms) on the
surface, and that the concentration of this adsorbate population can be
controlled to some extent.\cite{erwin_phys_rev_b_2009a} No such studies have yet been reported
for Si(553)-Au or Si(557)-Au, but the possibility of
tuning surface magnetism using surface chemistry suggests a
wealth of new research possibilities.

Linear chains of spin-polarized atoms provide atomically perfect
templates for the ultimate memory and logic, in which a
single spin represents a bit.
Here the spins are locked into
a self-assembled rigid lattice on a substrate that is
compatible with silicon technology. One potential application is the
spin shift register recently proposed theoretically 
by Mahan.{\cite{mahan_phys_rev_lett_2009a} 
This device requires a one-dimensional chain of identical atoms each
with spin $S=1/2$.  When an additional electron is conducted along
the chain, each spin state is shifted by one atom---the spin
analogue of a standard shift register memory device. The arrangment
of spins is arbitrary and there is no requirement of long-range
order.  However, two other important criteria must be satisfied: in
the ground state each atom must have one electron, and the atoms
must have correlated electronic states.  The first criterion is
indeed satisfied by the spin-polarized atoms on Si(553)-Au and
Si(557)-Au. In previous work we showed that the electronic states of
singly-occupied silicon dangling bonds are indeed strongly
correlated.\cite{hellberg_phys_rev_lett_1999a}

Another potential application is the storage of information in single
magnetic atoms.\cite{otte_europhysics_news_2008a} Toward this end,
Hirjibehedin {\em et al.}~recently used spin excitation spectroscopy
to demonstrate that the orientation of an individual spin can be
measured with an STM.\cite{hirjibehedin_science_2007a} For information
storage the stability of this orientation, for example from coupling
of the spin to the substrate, is critical. This coupling is most
likely to be substantial when the substrate is anisotropic at the
atomic scale. In the experiments of Hirjibehedin this anisotropy was
engineered by creating Cu$_2$N islands that break the fourfold
symmetry of the Cu(001) substrate. In the Si(553)-Au and related
vicinal silicon surfaces this anisotropy arises naturally at the step edge.
The investigation of spin-lattice coupling in these magnetic silicon
surfaces is an important next step toward the ultimate goal of
spintronics with single spins.

\vspace*{32pt}

\begin{flushleft}{\bf {\large Methods}}\end{flushleft}
First-principles total-energy calculations were used to determine
equilibrium geometries and relative energies of the models in Fig.~1
and their variants.  The calculations were performed in a slab
geometry with six or more layers of Si plus the reconstructed top
surface layer and a vacuum region of at least 10 \AA. All atomic
positions were relaxed, except the bottom Si layer and its passivating
hydrogen layer, until the largest force component on every atom was
below 0.02 eV/\AA. Total energies and forces were calculated within
the PBE generalized-gradient approximation
\cite{perdew_phys_rev_lett_1996a} to density-functional theory (DFT)
using projector-augmented-wave potentials, as implemented in {\sc
vasp}
\cite{kresse_phys_rev_b_1993a,kresse_phys_rev_b_1996a}.  The
plane-wave cutoff for all calculations was 250 eV.

The sampling of the surface Brillouin zone was chosen according to the
size of the surface unit cell and the relevant precision requirements.
For the 1$\times$6 reconstruction of Si(553)-Au shown in Fig.~1(a,b),
the equilibrium geometries and total energies of nonmagnetic,
ferromagnetic, and antiferromagnetic configurations were calculated
using 6$\times$6 zone sampling, with convergence checks using
10$\times$10 sampling. For the 1$\times$2 reconstruction of Si(557)-Au
shown in Fig.~1c, the equilibrium geometries and total energies of
nonmagnetic, ferromagnetic, and antiferromagnetic configurations were
calculated using 2$\times$8 zone sampling, with convergence checks using
2$\times$16 sampling.

Exchange coupling constants $J_{\parallel}$ and $J_{\perp}$ for
Si(553)-Au were extracted from DFT total energies. Supercells
containing two spins on each of two adjacent steps were constructed
with the four possible unique spin configurations. The DFT total
energies were fit to a nearest-neighbor classical Heisenberg
Hamiltonian. The numerical reliability of the coupling constants was
analyzed with respect to Brillouin-zone sampling, plane-wave cutoff,
and relativistic treatment, and the dependence of the coupling
constants upon charge doping of the surface states was investigated
(see Supplementary Information).

The simulated STM images in Fig.~2 were calculated using the
method of Tersoff and Hamann \cite{tersoff_phys_rev_b_1985a}. For this
filled-state image we integrated the local density of states (LDOS)
over a 0.5-eV energy window of occupied states up to the Fermi level.
The simulated STM topography under constant-current conditions was
obtained by plotting the height at which the integrated LDOS is
constant.


\begin{flushleft}{\bf {\large Acknowledgements}}\end{flushleft}
Helpful conversations with David L. Huber 
are gratefully acknowledged. This work was supported by the
Office of Naval Research, and by the NSF under awards No. DMR-0705145
and DMR-0084402 (SRC). Computations were performed at the DoD Major
Shared Resource Center at AFRL.

\end{document}